\documentclass[conference]{IEEEtran}
\IEEEoverridecommandlockouts
\usepackage{cite}
\usepackage{amsmath,amssymb,amsfonts}
\usepackage{algorithm}
\usepackage{graphicx}
\usepackage{textcomp}
\usepackage{xcolor}
\usepackage{algpseudocode}
\usepackage{url}
\usepackage{float}

\usepackage{color,soul}
\usepackage{tikz}
\def\BibTeX{{\rm B\kern-.05em{\sc i\kern-.025em b}\kern-.08em
    T\kern-.1667em\lower.7ex\hbox{E}\kern-.125emX}}
\begin{document}

\title{Multi-Layer Monitoring at the Edge for Vehicular Video Streaming: Field Trials}
% \\

\author{\IEEEauthorblockN{Inhar Yeregui, Juncal Uriol\IEEEauthorrefmark{1}, Roberto Viola}
\IEEEauthorblockA{\textit{Fundaci\'on Vicomtech} \\
\textit{Basque Research and Technology Alliance}\\
San Sebasti\'an, 20009 Spain\\
\{iyeregui, juriol, rviola\}@vicomtech.org}
\IEEEauthorrefmark{1}PhD Candidate at UPV/EHU
%\\
\and
\IEEEauthorblockN{Pablo Angueira, Jasone Astorga, Jon Montalb\'an}
\IEEEauthorblockA{\textit{Department of Communications Engineering} \\
\textit{University of the Basque Country (UPV/EHU)}\\
Bilbao, 48013 Spain\\
\{pablo.angueira, jasone.astorga, jon.montalban\}@ehu.eus}

}

\maketitle

\IEEEoverridecommandlockouts
\IEEEpubid{\begin{minipage}{\textwidth}\ \\\\\\\\\\[12pt]\centering
I. Yeregui et al., "Multi-Layer Monitoring at the Edge for Vehicular Video Streaming: Field Trials", 2023 IEEE International Symposium on Broadband Multimedia Systems and Broadcasting (BMSB), 2023, pp. 1-7, doi: 10.1109/BMSB58369.2023.10211216. \copyright 2023 IEEE. Personal use of this material is permitted. Permission from IEEE must be obtained for all other uses, in any current or future media, including reprinting/republishing this material for advertising or promotional purposes, creating new collective works, for resale or redistribution to servers or lists, or reuse of any copyrighted component of this work in other works.
\end{minipage}}

\begin{abstract}
In an increasingly connected world, wireless networks' monitoring and characterization are of vital importance. Service and application providers need to have a detailed understanding of network performance to offer new solutions tailored to the needs of today's society. In the context of mobility, in-vehicle infotainment services are expected to stand out among other popular connected vehicle services, so it is essential that communication networks are able to satisfy the Quality of Service (QoS) and Quality of Experience (QoE) requirements needed for these type of services. 
This paper investigates a multi-layer network performance monitoring architecture at the edge providing QoS, QoE, and localization information for vehicular video streaming applications in real-time over 5G networks.
In order to conduct field trials and show test results, Mobile Network Operators (MNOs)' 5G Standalone (SA) network and Multi-access Edge Computing (MEC) infrastructure are used to provide connectivity and edge computing resources to a vehicle equipped with a 5G modem.
\end{abstract}


\section{Introduction}
The automotive industry is moving rapidly towards the commercialization of connected and autonomous vehicles.
%and on this path the achievement of the connected vehicle is an important milestone.
In the process of achieving this goal, vehicle location services and fifth generation (5G) cellular networks play an important role, mainly as enablers of vehicle-to-vehicle (V2V), or vehicle-to-infrastructure (V2I) communications \cite{houmer2022lightweight}. These communications are essential for vehicles to transmit information to each other and for infrastructures to manage these communications in a coordinated manner. Thus, they enable several use cases, such as Smart Parking \cite{al2019smart} and Truck Platooning \cite{balador2022survey}. 

In parallel, video streaming applications are gaining popularity \cite{Netzorg2021PopFactorLB}, becoming the major source of Internet traffic, and their usage is constantly growing. According to forecasts, video is expected to account for 80 percent of global mobile network traffic in 2028 \cite{ericsson}. The inevitable merging of these two realities causes the emergence of in-vehicle infotainment services \cite{jiang2021resource}, and their upward trend is expected to be maintained in the next years \cite{systems10050162}. To support this trend, the communication channels need to satisfy the Key Performance Indicators (KPIs) of this kind of services. This makes channel characterization and service monitoring essential for the development of the mentioned services.

5G technologies promise improved performance in terms of increased throughput, reduced latency, and increased reliability under high mobility and user-density environments \cite{giust2018multi}. New advanced technological features, such as virtualization, softwarization, or network slicing, will be the key to achieving those goals.
%This set of improvements makes 5G networks suitable for several use cases that require markedly reliable and fast connections, such as Smart Parking \cite{al2019smart}, Truck Platooning \cite{balador2022survey} or vehicular video streaming \cite{jiang2021resource}. 

In particular, Multi-access Edge Computing (MEC), a new network architecture concept under the 5G umbrella, enables cloud computing capabilities and an IT service environment at the edge of the network. Its privileged position, closer to the end users, allows for reduced latency, ensures highly efficient network operation and service delivery, and improves the customer experience \cite{9240934}. Moreover, it enables the monitorization of network traffic exchanged between the core network and the RAN, allowing the optimization of the operations of any service running on the network.

This paper investigates a multi-layer network performance monitoring architecture at the edge providing real-time QoS, QoE, and localization information for vehicular video streaming applications over 5G networks.
The solution is achieved by providing the following relevant contributions:

\begin{itemize}

    \item Design of a modular
    % and virtualized
    %client/server model
    architecture
    %for the network edge
    to monitor multiple layers of the Open Systems Interconnection (OSI) model \cite{zimmermann1980osi}. The architecture is composed of different containerized services that can be easily deployed on top of any virtualized host.
    
    \item Implementation of the proposed architecture on top of MNOs' 5G SA and MEC-enabled network
    %, including 5G Radio Access Network (RAN) and Multi-access Edge Computing (MEC) infrastructure used
    employed to deliver multimedia streams.
    %to connected vehicles.
    
    \item Validation of the proposed solution showing field trials and test results based on a real
    %cooperative, connected, and automated mobility (CCAM)
    mobility scenario, where a Dynamic Adaptive Streaming over HTTP (DASH) stream is
    %transmitted through a 5G Standalone (SA) network to a vehicle in motion.
    transmitted over the network and received by a player run in a vehicle.
    
\end{itemize}

The rest of the paper is structured as follows.
Section \ref{sec:RelatedWork} reviews the related work in the domain of multi-layer monitoring solutions applied to vehicular video streaming.
%media delivery through telecommunication networks.
In section \ref{sec:CLMA}, our modular architecture for multi-layer monitoring is described, and the advantages of using MEC capabilities for its deployment are explained.
Section \ref{sec:implementation} and Section \ref{sec:results} present the implementation of the solution over a real 5G setup and the experimental assessment of the proposed solution, based on field trials, respectively.
%The setup comprises 5G RAN and MEC infrastructure and a Dynamic Adaptive Streaming over HTTP (DASH) streaming service.
% Section \ref{sec:results} presents the results of the field trials.
%obtained by testing the solution in a real
%Cooperative, connected and automated mobility (CCAM)
%mobility scenario where a DASH player, running in a connected vehicle
%has been used in order to consume
%and consuming a video streaming service, has been monitored.
Finally, we present our conclusions and future work in Section \ref{sec:conclusions}.

\section{Related Work}
\label{sec:RelatedWork}
% Network and service performance monitoring is an interest topic for the past few years due to their contribution to the development and enhancement of their own capabilities.
In recent years, network and service performance monitoring has been a rising topic, enabling the development of more intelligent services. A service operates depending on target KPIs and adjusts its operations depending on the network workload at any time. Thus, the network is constantly monitored to manage the life-cycle of virtual functions, detect network issues or QoS violations and perform actions that restore the proper operation.

%This section reviews the related work based on multi-layer monitoring.

%According to ETSI specification, network monitoring tasks can be passive, active, or hybrid \cite{etsi2016monitoring}. 
%Passive monitoring consists of observing network traffic generated by network applications and users. It is limited to collecting data already available at the network agents. Here, the type of traffic and the duration of network flows influence the measurements \cite{yu2013flowsense}. However, measurements may not be available during periods when network traffic is not generated. 

%On the contrary, active measurements aim to perform a more extensive diagnostic of the network conditions and determine if network packets are correctly transferred between hosts. For that, it generates and sends bulk data flows to analyze the behavior of the network \cite{hofstede2014flow}. Finally, hybrid network monitoring combines passive and active measurements to extract a deeper overview of the network status. 

%\hl{Another important aspect... -> Network performance under multimedia scenarios}

%Active monitoring involves the generation of synthetic or test traffic to validate network applications' performance and verify that the SLA is fulfilled. Then, instead of simply collecting information available at network agents, it generates and sends traffic flows to analyze the behaviour of the network \cite{hofstede2014flow}. 

%Several tools are proposed in the literature to accomplish the monitoring task and to gather information on different kinds of metrics. 

Several tools are proposed in the literature to accomplish the monitoring task and gather information on different metrics. Multiple layers of the OSI model can be considered when carrying out monitorization tasks. Some works focus their research on network and channel characterization for different generations of mobile networks, combining physical layer (L1) metrics such as Reference Signal Received Power (RSRP), Reference Signal Received Quality (RSRQ), and Signal Interference Noise Ratio (SINR) and network layer (L3) metrics such as throughput and round-trip time (RTT). Others center the efforts on monitoring metrics from the transport layer (L4) \cite{lopez2018virtualized} or even monitoring the network performance under multimedia streaming scenarios collecting application layer metrics (L7) such as video bitrate.

In the work presented in \cite{fernandez2021railway}, an innovative tool called Channel Characterization Tool (CCT) is presented in order to collect physical and network layer metrics to fulfill the purpose of the railway migration task. It is customized for collecting metrics in railway environments. Therefore, the collected metrics are more accurate along a track than other applications.
%as it is presented in \cite{tan2007empirical}, \cite{fernandez2021railway}
%\cite{tan2007empirical}, \cite{fernandez2021railway}. -> L3 metrics

%, and some physical aspects  \cite{bokani2016comprehensive} 

%\hl{Monitoring of network L1/L2/L3 -> TFM de Inhar}

%MEC is a type of network architecture that provides cloud computing capabilities and an IT service environment at the edge of the network. The goal of MEC is to reduce latency, ensure highly efficient network operation and service delivery, and improve the customer experience \cite{9240934}. It is in a privileged position to monitor the traffic that is exchanged between the core network and the RAN.

%\hl{L1/2/3 Monitoring}
% \cite{fernandez2021railway}

Concerning media-specific monitoring in \cite{viola2022multi}, a MEC proxy is proposed for monitoring all the traffic exchanged at RAN between the video players, media server, and Content Delivery Network (CDN). That MEC proxy collects metrics of the application layer (L7) in relation to the streaming session. 

Following with application layer monitoring, in \cite{laiche2021qoe}, the authors present a QoE monitoring system called WebQoE, which introduces a system for collecting video streaming performance and quality metrics. This monitoring system is a web application replicating a video streaming service to measure its performance. Then, users must rate their experience to collect quality metrics. Therefore, it is a measuring tool for collecting objective and subjective metrics.

Moreover, in \cite{mangla2017mimic}, authors present a methodology for estimating video QoE metrics. They collect passive measurements using HTTP logs from the network and use them for estimating QoE metrics. This work is extended in \cite{mangla2018emimic}, in which they reconstruct the methodology previously proposed for estimating QoE metrics with Encrypted Network Traffic.

%\cite{laiche2021qoe}

%\hl{QoE}
%Mangla T, Halepovic E, Ammar M, Zegura E (2017) MIMIC: Using passive network measurements to estimate HTTP-based adaptive video QoE metrics. In: 2017 network traffic measurement and analysis conference, pp 1–6. https://doi.org/10.23919/TMA.2017.8002920
%20. Mangla T, Halepovic E, Ammar M, Zegura E (2018) emimic: Estimating http-based video qoe metrics from encrypted network traffic. In: 2018 network traffic measurement and analysis conference, pp 1–8. https://doi.org/10.23919/TMA.2018.8506519

%\hl{Solutions that do both network and media monitoring -> are there any in literature? (Inhar found a paper for multi layer)} \cite{LOPES2022100534}

% Introduccion a multi-layer
In relation to multi-layer monitoring, in \cite{LOPES2022100534}, the authors present a multi-layer probing mechanism for collecting network metrics among several layers. They use a Customer Premises Equipment (CPE), like a proxy between the server and the client, as an intermediate node that grants the client connection to outside. This CPE consists of different components that recollect physical, network, transport, and application layer metrics. 

%: performance, quality, and experience modules. First, the performance module recollects physical and network metrics (L1-L3). Then, the quality module collects information extracted from transport layer communication (L4). Finally, the experience module allows estimating metrics directly from video characteristics (L5).

%\subsection{Multi-Access Edge Computing}

%\hl{Differently from previous works, our solution proposes an architecture capable of monitoring a 5G network at the edge using a MEC for video streaming combining metrics from different OSI layers such as physical layer (L1), network layer (L3) and application layer (L7). This monitoring architecture allows the collection of QoS, QoE and localization information in order to have a major knowledge of the 5G network in vehicular use cases.}

%All of these works contribute to orienting our work proposal by focusing on keeping the strengths of the different monitoring solutions, such as supplementing the network characterization with media-specific metric monitoring and enhancing additional aspects related to monitoring methods, such as the involvement of the MEC.

The main contribution of this paper is the introduction of a multi-layer monitoring system at the edge, which is capable of extracting metrics at different OSI layers for vehicular video streaming applications. It has been tested using MNOs' 5G network and MEC infrastructure. The monitored system combines physical (L1), network (L3), and application (L7) layer metrics. Therefore, the presented monitoring architecture allows the collection of QoS, QoE, and localization information in order to have a major knowledge of the 5G network in vehicular use cases.

%using  different OSI layer metrics: L1, L2, L3 and L7.

\section{Multi-layer monitoring architecture}
\label{sec:CLMA}

\begin{figure*}[!htbp]
\centerline{\includegraphics[width=0.95\textwidth,keepaspectratio]{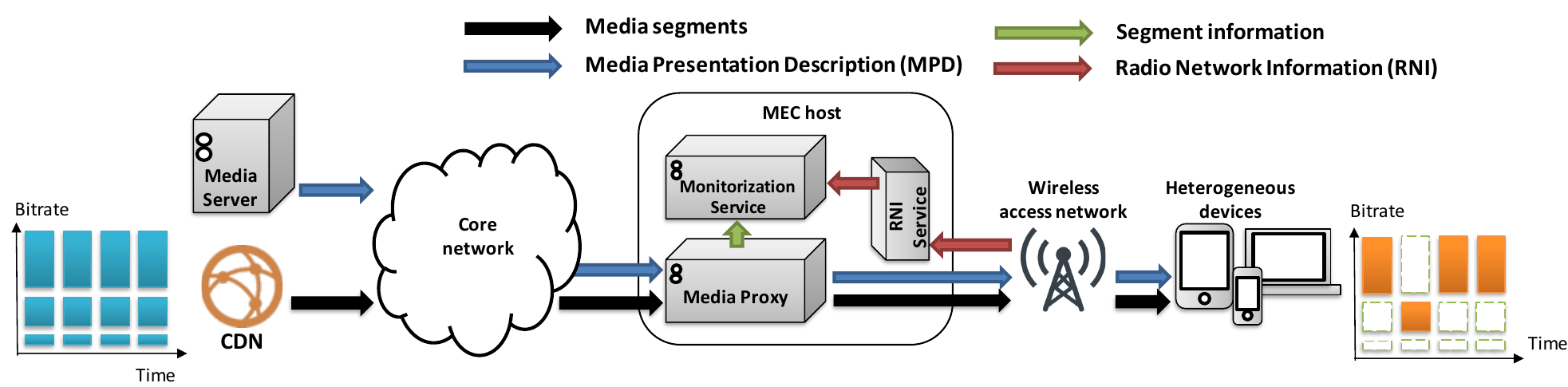}}
\caption{General architecture of the solution.}
\label{fig:architecture}
\end{figure*}

% The analysis of both QoS and QoE is crucial for decision making system enforcing or boosting multimedia streaming services. This section describes the general architecture proposed to fulfill the purpose of this paper. 

% As Figure \ref{fig:architecture} shows, the designed solution follows a server-client architecture, where the client consumes a video streaming service that is located on a server. One of the most noteworthy features of the solution is the deployment of edge capabilities...

% Another strength of this monitoring solution is that it is able to gather metrics from multiple layers of a telecommunication network. Having knowledge of the transmission channel's health and capabilities is of vital importance when developing any application or service that is going to run on top of it. The main KPIs for CCAM applications are related to layers 1, 3 and 7, which correspond to the physic layer, the network layer and the application layer \cite{santa2022evaluation}. That is why the metrics that this monitorization tool obtains are the following:

The general architecture proposed to perform multi-layer monitoring of a video streaming service is presented in Figure \ref{fig:architecture}. As the solution aims at collecting measurements at different OSI layers, the monitoring system needs to access information on gNodeB-modem connectivity (L1/3) and running player status (L7).

When considering gNodeB-modem connectivity, information on lower ISO layers can be achieved at the MEC, as ETSI defines a Radio Network Information Service (RNI Service or RNIS) \cite{etsi2017mobile}. RNIS is responsible for interacting with the Radio Access Network (RAN), collecting RAN-level information about User Equipment (UE), and exposing it to any edge application through a dedicated RNI Application Programming Interface (API). The application can use the API and the provided information to dynamically adjust its behavior to optimally match the RAN conditions \cite{giust2017multi}. Moreover, the involvement of MEC capabilities in a monitoring system deployment provides the possibility to enhance the solution's scalability, maintaining a local vision of the tested scenario, and exploiting the collected data to achieve more sophisticated automotive services \cite{avino2019mec}. Edge monitoring is essential for virtual resources such as containers that serve microservices from the edge, which may need to be dynamically spun up or down as needed. To respond in real-time to rapidly changing traffic and device density with on-demand virtual infrastructure, mobile operators must employ orchestration. And to achieve this level of automation, orchestrators need real-time smart data prepared and organized at the collection point so it is ready and optimized for analytics at the highest quality and speed to inform it.

When considering the player, its status is representative of application layer information (L7). Specifically, when considering DASH streams, the player uses HTTP protocol to receive the video content. It means that the design follows a typical HTTP server-client architecture, where the DASH player retrieves the Media Presentation Description (MPD) at the Media Server and then accesses the DASH media segments thanks to \textit{BaseURL} information contained in the MPD.

In a legacy scenario, the \textit{BaseURL} addresses the player to download the media segments from the CDN. On the contrary, when an intermediate MEC node is introduced, the \textit{BaseURL} can address the MEC where a media proxy service is in charge of receiving the HTTP requests from the player and providing the response by retrieving the content from the CDN. The introduction of a media proxy opens the possibility of tracking a player's activity and collecting information on its playback session. Moreover, collecting information from MPD and media segments allows for estimating the QoE of each user \cite{viola2022multi} .

%\begin{figure}[t!]
%\centerline{\includegraphics[width=0.35\textwidth,keepaspectratio]{modules.pdf}}
%\caption{Modules included in for the monitorization service.}
%\label{fig:modules}
%\end{figure}

The information provided by both RNIS and media proxy is stored in a monitorization service at the MEC host for real-time visualization or exploitation by any other service. 
%In Figure \ref{fig:modules}, the different modules provided by the monitorization service are shown.
The monitorization service consists of four key modules: the Radio Network data collector, the Media Session data collector, the QoE analyzer and the Data visualizer.
Radio Network data collector and Media Session data collector modules are responsible for retrieving L1/3 metrics and player L7 information, respectively. The QoE analyzer module executes the ITU-T P.1203 model \cite{itup1203}, in order to obtain QoE-related metrics from the L7 information. Finally, the Data visualizer module offers a visual analysis service, showing all the previously mentioned data.
All the modules can be deployed as separate containerized services on top of a common virtualized MEC host. Typically, vehicular scenarios tend to observe a variable demand due to the mobility of the UEs. Thus, virtualized and modular solutions satisfy that need for mobility as containers are suitable for dynamic deployment, scale or migration operations \cite{khan2019conceptual}.

\begin{figure}[t!]
\centerline{\includegraphics[width=0.4\textwidth,keepaspectratio]{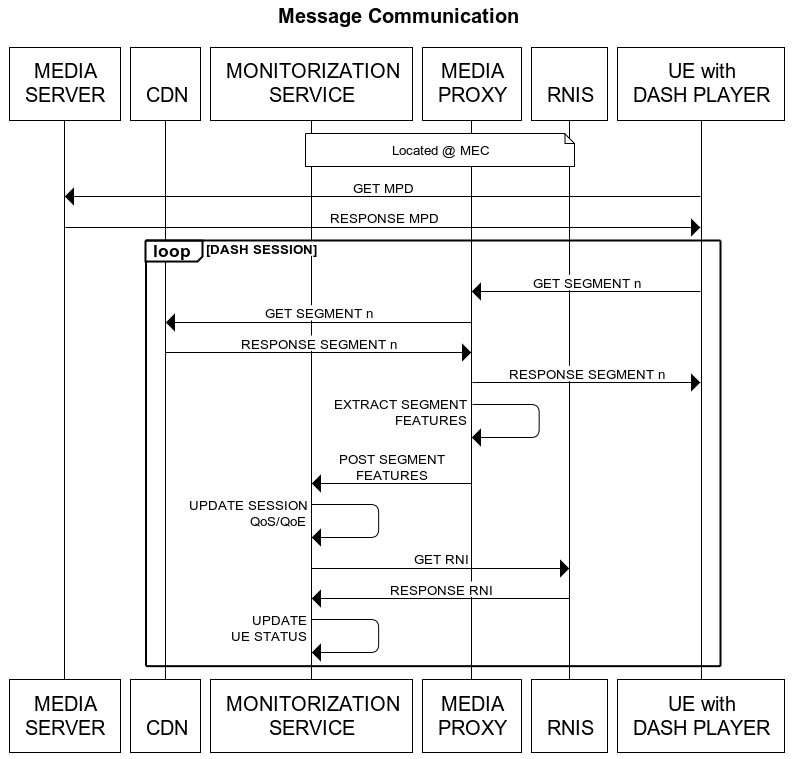}}
\caption{Message communication.}
\label{fig:comms}
\end{figure}

% When combining the information of radio access and the player status in a unique process,
The general communication flow to store multi-layer information in the monitorization service is presented in Figure \ref{fig:comms}. Every time a media segment is requested by the player, the media proxy retrieves the segment from the CDN, extracts the segment information to feed the monitorization service, and then, serves it to the player. The monitorization service uses the segment information to estimate the QoE. In parallel, it also retrieves RNI through the RNIS, so it can track and monitor a service or application performance considering multiple communication layers.

% When considering QoE model, ITU-T P.1203 is the standard for DASH. Further details are presented in \cite{viola2022multi}.

% \begin{table}[htbp]
% \caption{Table Type Styles}
% \begin{center}
% \begin{tabular}{|c|c|c|c|}
% \hline
% \textbf{Table}&\multicolumn{3}{|c|}{\textbf{Table Column Head}} \\
% \cline{2-4} 
% \textbf{Head} & \textbf{\textit{Table column subhead}}& \textbf{\textit{Subhead}}& \textbf{\textit{Subhead}} \\
% \hline
% copy& More table copy$^{\mathrm{a}}$& &  \\
% \hline
% \multicolumn{4}{l}{$^{\mathrm{a}}$Sample of a Table footnote.}
% \end{tabular}
% \label{tab1}
% \end{center}
% \end{table}

% \begin{figure}[htbp]
% \centerline{\includegraphics{fig1.png}}
% \caption{Example of a figure caption.}
% \label{fig}
% \end{figure}

\section{Implementation}
\label{sec:implementation}
For the implementation of the proposed multi-layer monitoring solution, different software tools have been integrated. 
Two different MNOs have collaborated to provide the telecommunication infrastructure where the solution has been deployed: Euskaltel provided the 5G Core and MEC infrastructure, while Orange provided the RAN.
In the absence of a functional RNIS at the MEC, a different approach has been sought in order to gather data related to multiple layers of the communication stack. The proposed solution consigns the responsibility of collecting RAN information to the UE. Then, this information is transferred to the Monitorization Service located at the MEC Host.

\begin{figure}[t!]
\centerline{\includegraphics[width=0.5\textwidth,keepaspectratio]{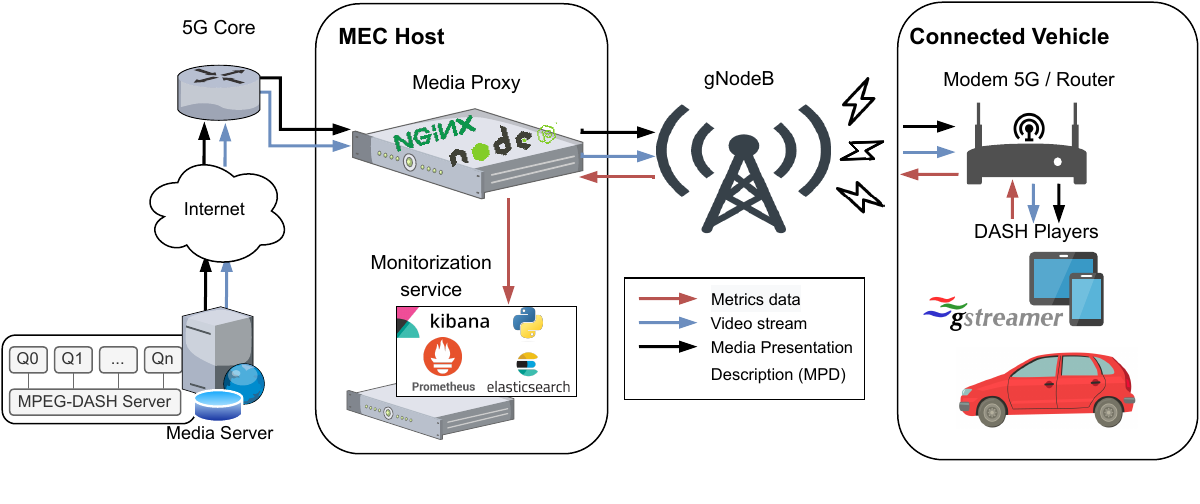}}
\caption{Low-level architecture scheme of the proposed solution.}
\label{fig:lowlevel}
\end{figure}

% The following is the description of the different main blocks that comprise the final setup and that are shown in Figure \ref{fig:lowlevel}:
Figure \ref{fig:lowlevel} shows the main blocks that comprise the final setup:
\begin{itemize}
    
    \item UE with DASH Player: a DASH player based on GStreamer multimedia framework \cite{gstreamer} is executed in a 5G-connected UE located in the connected vehicle. This UE has also been provided with an L1 and L3 metric
    %collector, which consists of a
    exporter based on Python3 that communicates with the 5G modem in order to retrieve
    Radio Frequency (RF) metrics (L1) and Global Positioning System (GPS) information (via AT commands) and monitor the corresponding network interface (L3) to obtain traffic-related metrics.
    Then, the exporter exposes the collected measurements to the Monitorization service.
    The employed 5G modem is a Telit FN980.
    % This way, the UE will consist of a MPEG-DASH player with a dedicated service collecting multi-layer data metrics.
    
    \item 5G Core, MEC Host and gNodeB: Euskaltel MNO's 5G Core network and virtualized MEC infrastructure and Orange MNO's 5G base station.
    % the 5G network base station, which allows the wireless UEs, such as 5G modems or smartphones, to connect to the internet.

    % \item 5G Core and MEC Host: 5G Core network and virtualized MEC infrastructure provided by Euskaltel MNO.
    
    \item Media Proxy: a containerized proxy node based on Node.js \cite{nodejs} and NGINX \cite{reese2008nginx}, located at the MEC. It retrieves the DASH segments from the media server and forwards them to the player. 
    % It describes a model for monitoring media session quality while delivering content through HAS technologies. It receives media stream information and playback device features and generates scores for audio-visual and buffering. Finally, it provides the overall score which represents the Quality of Experience. All the outputs have 1-5 quality scale, where “1” means “bad” quality and “5” means “excellent” quality, according to MOS specifications.

    \item Monitorization Service: a node at the MEC including several containers. Each container implements a different module to collect and visualize data.
    A Prometheus \cite{schunke2007prometheus} module pulls and stores all the metrics coming from the players. In parallel, an Elasticsearch \cite{gormley2015elasticsearch} module collects the metrics coming from the Media Proxy and merges them with the ones in Prometheus.
    A Python-based implementation of the ITU-T P.1203 recommendation \cite{Robitza2018} is employed as a module for estimating QoE scores.
    A Kibana \cite{gupta2015kibana} module is deployed in order to visually analyze all the gathered metrics.
    
    \item Media Server: a server at ATHENA Christian Doppler Laboratory that publicly provides a multi-codec DASH dataset \cite{taraghi2022multi}. We choose the "Seconds that Count" video sequence as it is the longest available, with a duration of 322 seconds. Moreover, we select H265/HEVC encoded DASH stream, available at 15 different representations. The representations range goes from 320x180 at 145kbps to 7680x4320 (8K) at 27.5Mbps. The segment duration is set to 4 seconds.

\end{itemize}

\section{Experimental assessment}
\label{sec:results}

This section describes the results obtained by testing the proposed multi-layer monitoring solution in a mobility scenario, where a connected vehicle is provided with 5G SA connectivity and consumes a media service, namely, a DASH video stream. Tests have been carried out at Miramon technology park in San Sebasti\'{a}n, Spain. Table \ref{tab:network} describes specific parameters of the implemented RAN provided by the MNO. The road section traveled at the tests covers a diameter of approximately 1 km, and the furthest point from the antenna is 650 meters away. 

\begin{table}[htp]
\caption{RAN infrastructure specifications}
\centering
\def\arraystretch{1.2}%  1 is the default, change whatever you need
\setlength\tabcolsep{2.5pt}
\label{tab:network}
\begin{tabular}{|l|c|}
\hline
\textbf{Bandwidth}          & 100 MHz       \\ \hline
\textbf{Operation band}     & 3.5GHz        \\ \hline
\textbf{Transmission power} & 200 W         \\ \hline
\end{tabular}
\end{table}

Along the route, a video player has been repeatedly executed to play the dataset video sequence. In total, the video has been played 6 times, but only the first two player runs are selected to show the experimental assessment of the monitoring solution. For simplicity, the two player's runs are referred to as Player 1 and Player 2.

% The different figures shown in this section depict metrics that cover different layers of the OSI model: in a physical layer showcasing values of RSRP, RSRQ and SINR; in a network layer showing the received throughput and bytes caused by the video streaming application, and in an application level, considering received throughput, selected bitrate, latency, Stalls information and QoE values obtained by ITU-T P.1203 estimation algorithm. 

\begin{figure}[t!]
  \centering
  \begin{minipage}[b]{0.23\textwidth}
    \includegraphics[width=\textwidth]{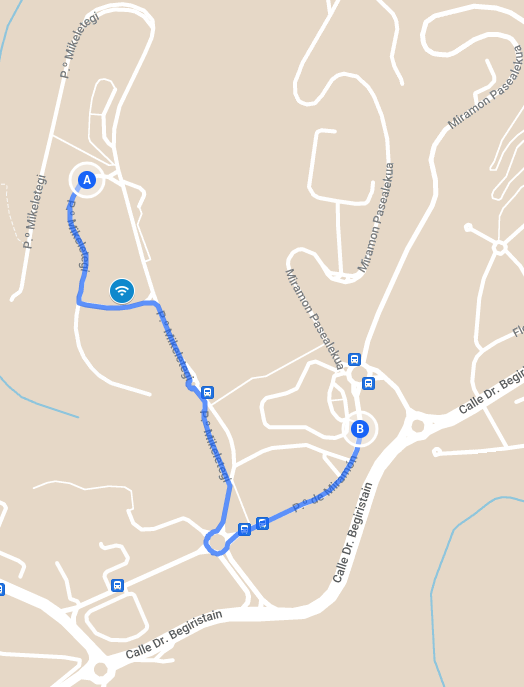}
  \end{minipage}
  \hfill
  \begin{minipage}[b]{0.23\textwidth}
    \includegraphics[width=\textwidth]{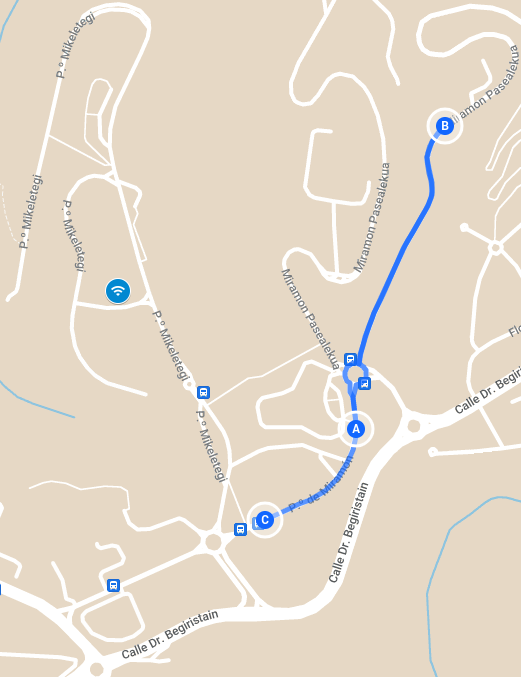}
  \end{minipage}
  \caption{Player 1 and Player 2 test maps.}
  \label{fig:player_maps}
\end{figure} 

\begin{figure}[t!]
\centerline{\includegraphics[width=0.5\textwidth,keepaspectratio]{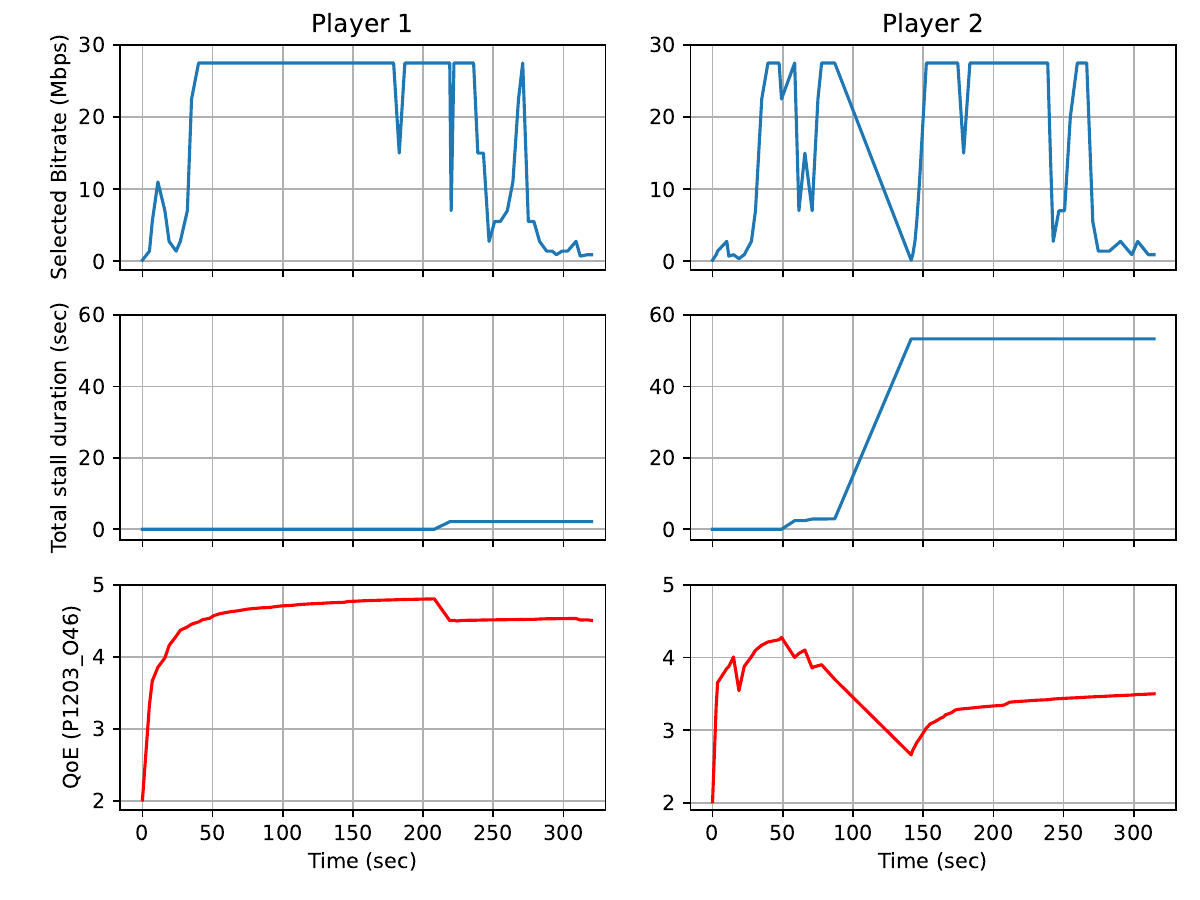}}
\caption{Selected Bitrate (Mbps), Total Stall Duration (sec), and QoE results for Player 1 and Player 2.}
\label{fig:l7}
\end{figure}

\begin{figure}[t!]
\centerline{\includegraphics[width=0.5\textwidth,keepaspectratio]{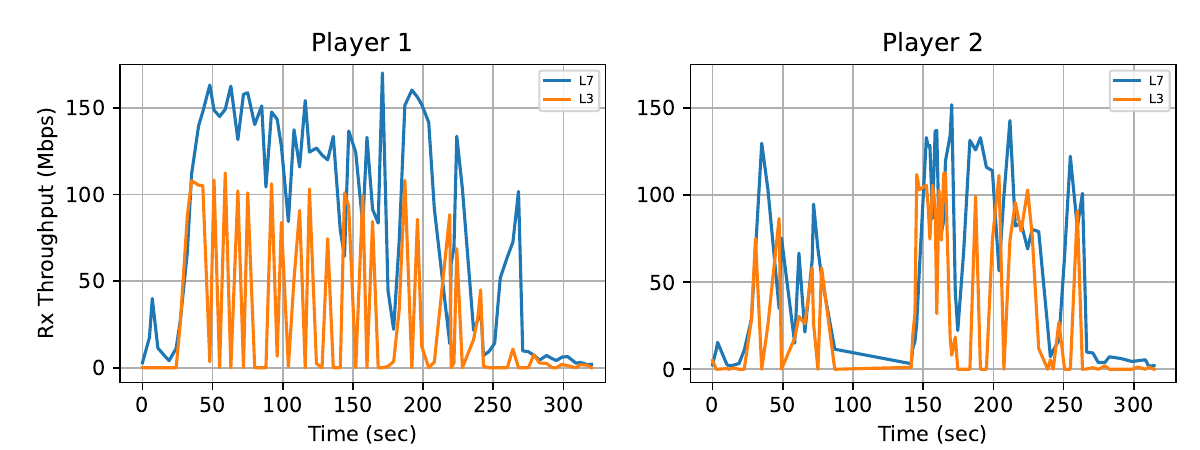}}
\caption{Comparison of L3/7 Throughput results for Player 1 and Player 2.}
\label{fig:multilayer}
\end{figure}

\begin{figure}[t!]
\centerline{\includegraphics[width=0.5\textwidth,keepaspectratio]{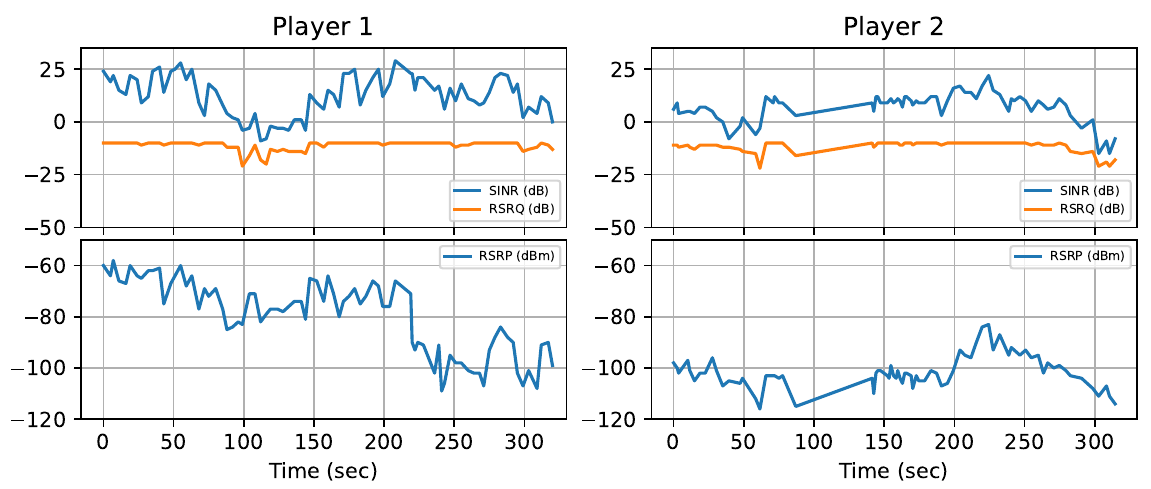}}
\caption{Achieved RF results for Player 1 and Player 2.}
\label{fig:rf}
\end{figure}

Figure \ref{fig:player_maps} shows the paths traveled by Player 1 and Player 2 while they were playing the video. Each path is represented by a blue colored line. In the case of Player 1, the path starts at point "A" and finishes at point "B". In the case of Player 2, the path starts at point "A" and ends at point "C", passing through point "B". Both images show the position of the antenna, and it is obvious that Player 1 has been executed in a more favorable area than Player 2, as it is closer to the antenna, so it is expected to obtain better QoS and QoE values than Player 2. This is clearly reflected in the obtained results, which can be seen in Figures \ref{fig:l7}, \ref{fig:rf}, and \ref{fig:multilayer}. 

%Player 1 was executed mostly in excellent and good signal zones, while Player 2 was executed in mid-cell and cell edge areas, where the signal strength is low. This is clearly reflected in the obtained results, which can be seen in Figures \ref{fig:rf}, \ref{fig:multilayer}, and \ref{fig:l7}. 
% {Moreover, Player 2's route shows that it moves away from the gNodeB until it reaches the furthest point, and then it comes back, so it can be expected that the QoS and QoE values get worse at first, but then they get better}.

Figure \ref{fig:l7} shows different L7 metrics obtained during the DASH streaming sessions, such as the selected video bitrate, representative of the quality of each DASH media segment downloaded by the players, the total stall duration that the players have experienced during the playback, and the evolution of the QoE score, obtained with the ITU-T P.1203 model. As expected, the selected bitrate drops as the vehicle moves away from the antenna (Figure \ref{fig:player_maps}). In the case of Player 1, it starts at 27.5 Mbps, which is the maximum value, but during the second half of the execution, it gets unstable and fluctuant. In the case of Player 2, the instability remains during the entire run, reaching 0 Mbps at a certain point. 
%, and the network throughput gets lower (Figure \ref{fig:multilayer}). 
In the same way, stalls tend to occur when the network conditions get worse, and the upward spikes shown in Total stall duration charts provide that information. Finally, the QoE is also affected, as ITU-T P.1203 model infers representation bitrate and stall information to estimate it. At first glance, the lower points of the curve concur with the furthest geographical points from the antenna, meaning worse network condition areas. Nevertheless, a multi-layer monitorization of the network, where different aspects of the communication stack are observed, enables the possibility of verifying it. 

% Selected Bitrate charts show the resolution of the segments that each player has downloaded and played. Player 1 (Figure \ref{fig:player1}g) is more stable and most of the segments have been downloaded at maximum bitrate, which is 27,5 Mbps. Player 2 (Figure \ref{fig:player1}h) suffers worse signal conditions, causing more instability and a lower mean value of selected bitrate during the download of the segments. 

% Total stall duration information is also monitored, as it can be a revealing factor when trying to evaluate different service KPIs. In the case of Player 1 (Figure \ref{fig:player1}i), signal power starts getting notably worse after the second 200 of the test, causing a stall of a couple of seconds. In the case of Player 2 (Figure \ref{fig:player1}j), the low signal power causes a stall of more than 50 seconds. 

% QoE data charts show the value that ITU-T P.1203 algorithm obtains. In the case of Player 1 (Figure \ref{fig:player1}k), QoE is almost 5, which is the highest value it can get, but after the second 200, it decreases a little bit. 
% This matches the behavior of the rest of the parameters. Looking at the map of the route, it can be deduced that at that moment the vehicle entered the orange/red zone.  In the case of Player 2 (Figure \ref{fig:player1}l), it just that the mean value for Player 1 is upper 4, while Player 2 experiences a value lower than 4 during almost the whole video.

Figure \ref{fig:multilayer} shows the download throughput of the players during the execution of both sessions. It compares the measurements collected at L3 and L7. By design, the measurements at L3 are performed every second, while at L7, they are performed each time a segment download finishes. It results that L7 measurements are not regular like L3 and have a lower frequency (a measurement for each segment duration, i.e., 4 seconds, approximately). Thus, L3 values also reveal the alternation of the download and idle states, typical of media segments download, while L7 values are taken only during the download state. Going deeper, this type of multi-layer monitoring makes it possible to compare and analyze the behavior of both curves, making it easier to understand different events that can happen in the communication channel. In this case, the peaks at L3 tend to occur before their respective ones at L7, enabling the possibility of exploiting higher frequency L3 measurements that are richer in information to forecast events at L7.

The charts shown in Figure \ref{fig:rf} describe the evolution of signal quality during both tests. They show RSRP, RSRQ, and SINR values over time. These are very significant L1 parameters when testing the performance of any wireless telecommunication network, as they represent the received signal power, quality, and signal-to-interference-plus-noise ratio. %These metrics provide information about the physical aspect of the network in specific zones inside the coverage area. 
In the case of Player 1, RSRP is declining due to the progressive loss of coverage, and between seconds 100 and 150, SINR experiences a noticeable decrease due to the existence of interference, causing the deterioration of QoE and QoS, as it can be seen in Figures \ref{fig:l7}, and \ref{fig:multilayer}.
% In the case of Player 2, the evolution of RSRP is more irregular and, in general, it is worse than for Player 1. In the case of SINR, it remains more stable.
In the case of Player 2, the evolution of RSRP shows an alternation between increasing and decreasing trends, and in general, it is worse than for Player 1. In the case of SINR, it remains more stable. The charts show that when measured RSRP approaches values below -100 dBm, the rest of QoS parameters worsen, and consequently, the QoE of the players is lower too.
The analysis of multiple layer metrics shows that the video streaming service's performance is directly related to the physical state of the network, as the L7 and L3 metrics show improvements when L2 metrics show a better state of the network, and the results deteriorate as L2 shows worse physical conditions. 

\begin{figure}[t!]
\centerline{\includegraphics[width=0.4\textwidth,keepaspectratio]{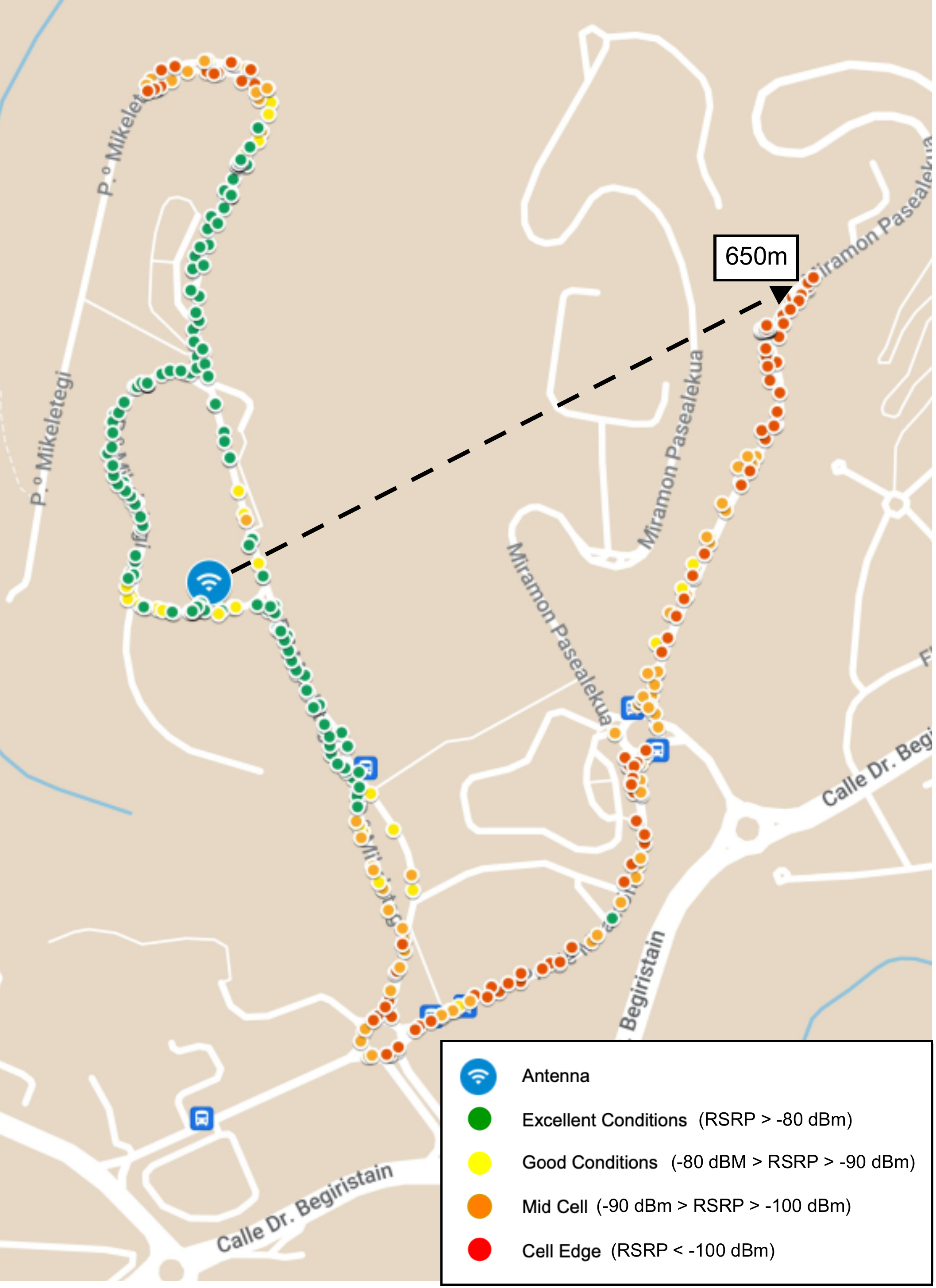}}
\caption{Coverage map of the track.}
\label{fig:coverage}
\end{figure}

In order to complete the service and network performance monitoring and characterization, these quality parameters have been supplemented with geolocalization metrics. This way, L2 metrics, along with the GPS coordinates obtained during the tests, have been used to draw a coverage map, which can be seen in Figure \ref{fig:coverage}. The green or excellent signal zone covers the area where the RSRP obtained is greater than -80 dBm. The yellow or good signal zone covers the area where the RSRP is between -80 and -90 dBm. Orange or mid-cell corresponds to the area where the RSRP is between -90 and -100 dBm, and finally, the red or cell edge area is the area where the RSRP is less than -100 dBm.

% Received Data charts show the amount of traffic received by the players, caused by the video streaming service. Comparing both player's charts it is obvious that Player 1 (Figure \ref{fig:player1}c) received all the video segments at a more regular speed, while Player 2 (Figure \ref{fig:player1}d) seems to have had more variation. Moreover, the number of mega bytes received by Player 2 in the same range of time is less than Player 1's case, meaning that Player 2 received lower resolution and lighter segments.

Gathering all these multiple OSI layer-related metrics enables having a very complete vision of the network's behavior and the video streaming service's performance assessed in a vehicular scenario. Moreover, being able to relate the obtained results to a specific location inside the coverage area makes it possible to assess different aspects of cell capabilities. 

% \begin{figure}[htbp]
% \centerline{\includegraphics[width=0.5\textwidth,keepaspectratio]{charts_1.pdf}}
% \caption{Achieved results for Player 1 (left column) and Player 2 (right column): RF Data (a) and (b), Rx Data (c) and (d), Rx Throughput (e) and (f), Selected Bitrate (g) and (h), Total stall duration (i) and (j), QoE (k) and (l).}
% \label{fig:player1}
% \end{figure}

% \end{figure}
% \begin{figure}[htbp]
% \centerline{\includegraphics[width=0.5\textwidth,keepaspectratio]{player1.png}}
% \caption{Player1 multi-layer data.}
% \label{fig:player1}
% \end{figure}

% \begin{figure}[htbp]
% \centerline{\includegraphics[width=0.5\textwidth,keepaspectratio]{player2.png}}
% \caption{Player2 multi-layer data.}
% \label{fig:player2}
% \end{figure}

% \begin{table}[H]
% \centering
% \caption{Video streaming stalls information}
% \label{tab:stalls}
% \begin{tabular}{l|c|c|}
% \cline{2-3}
%  & \multicolumn{1}{l|}{\textbf{Number of Stalls}} & \multicolumn{1}{l|}{\textbf{Totall Stall Duration (ms)}} \\ \hline
% \multicolumn{1}{|l|}{\textbf{Player 1}} & 1 & 2.131999969 \\ \hline
% \multicolumn{1}{|l|}{\textbf{Player 2}} & 4 & 13.333 \\ \hline
% \multicolumn{1}{|l|}{\textbf{Player 3}} & 1 & 4.930999994 \\ \hline
% \multicolumn{1}{|l|}{\textbf{Player 4}} & 5 & 4.123400021 \\ \hline
% \multicolumn{1}{|l|}{\textbf{Player 5}} & 1 & 1.249000072 \\ \hline
% \multicolumn{1}{|l|}{\textbf{Player 6}} & 3 & 2.059999943 \\ \hline
% \end{tabular}
% \end{table}

\section{Conclusions and Future Work}
\label{sec:conclusions}

This paper presented a multi-layer monitoring solution that implements edge capabilities to assess video transmission over 5G networks and shows field trials. The implemented solution has been tested in a mobility scenario, where a connected vehicle is provided with 5G SA connectivity and consumes a video streaming service. 

% by exploiting a dataset makes it very easy to relate application and network layer data with geo-localization and physical layer data. The different charts show the level of network workload at any time. This makes it possible to detect network issues and QoS violations that have an influence on the player's QoE.

The collection of multiple OSI layers-related metrics and the conduction of a streaming service characterization enables the possibility of having a detailed understanding of network performance. 

Moreover, the integration of MEC capabilities in the monitoring system deployment provides a real-time component to the solution, making it possible to enhance its scalability.
Thus, the collected data can be exploited in the future for decision-making algorithms to offer tailored solutions to the needs of today and future vehicular use cases.

% as input for decision making 
% Machine Learning models that can potentially improve the service performance.

% In the future, this work can be complemented with the deployment of enhanced vehicular infotainment services exploiting MEC capabilities, such as multi-layer metrics-based MEC caching instances, so that the content can be prefetched and served from the MEC host, reducing network traffic and increasing the performance.

\section*{Acknowledgment}

This research was supported by Red.es, Spain's 5G National Plan, under grant C012/12-SP for the 5G Euskadi project, and by Smart Networks and Services Joint Undertaking under the European Union’s Horizon Europe Research and Innovation programme, under Grant Agreement 101096838 for 6G-XR project.

% \section*{References}

% \end{thebibliography}

\bibliographystyle{IEEEtran}
\bibliography{main.bib}

\end{document}